\documentclass[reprint,superscriptaddress,amssymb,aps,prx,longbibliography]{revtex4-2}

\usepackage{amsmath}
\usepackage{mathtools}
\usepackage{xcolor}
\usepackage{hyperref}
\usepackage[T1]{fontenc}
\usepackage{amssymb}
\usepackage{graphicx}
\usepackage{ulem}

\begin{document}

\title{Experimental investigation of altermagnetic order in Cr-doped FeSb$_2$}

\author{A K M Ashiquzzaman Shawon}
\affiliation{Department of Physics, University of Michigan, Ann Arbor, MI 48109, USA}

\author{Eoghan Downey}
\affiliation{Department of Physics, University of Michigan, Ann Arbor, MI 48109, USA}

\author{Shane Smolenski}
\affiliation{Department of Physics, University of Michigan, Ann Arbor, MI 48109, USA}

\author{Thomas J. Hicken}
\affiliation{PSI Center for Neutron and Muon Sciences, 5232 Villigen PSI, Switzerland}

\author{Tatenda Kanyowa}
\affiliation{Neutron Scattering Division, Oak Ridge National Laboratory, Oak Ridge, Tennessee 37831, USA}

\author{Amir Henderson}
\affiliation{Department of Physics, University of Michigan, Ann Arbor, MI 48109, USA}

\author{Si Athena Chen}
\affiliation{Neutron Scattering Division, Oak Ridge National Laboratory, Oak Ridge, Tennessee 37831, USA}

\author{Mingyu Xu}
\affiliation{Department of Chemistry, Michigan State University, East Lansing, MI 48864, USA}

\author{Trisha Musall}
\affiliation{Department of Physics, University of Michigan, Ann Arbor, MI 48109, USA}

\author{Rafael Lopes Sabainsk}
\affiliation{Institute of Physics, Goethe-University Frankfurt, Max-von-Laue-Str. 1, 60438 Frankfurt am Main, Germany}

\author{Zachary J. Morgan}
\affiliation{Neutron Scattering Division, Oak Ridge National Laboratory, Oak Ridge, Tennessee 37831, USA}

\author{Wei Tian}
\affiliation{Neutron Scattering Division, Oak Ridge National Laboratory, Oak Ridge, Tennessee 37831, USA}

\author{Yuan Zhu}
\affiliation{Department of Physics, University of Michigan, Ann Arbor, MI 48109, USA}

\author{Weiwei Xie}
\affiliation{Department of Chemistry, Michigan State University, East Lansing, MI 48864, USA}

\author{Elena Gati}
\affiliation{Institute of Physics, Goethe-University Frankfurt, Max-von-Laue-Str. 1, 60438 Frankfurt am Main, Germany}

\author{Lu Li}
\affiliation{Department of Physics, University of Michigan, Ann Arbor, MI 48109, USA}

\author{Zurab Guguchia}
\affiliation{PSI Center for Neutron and Muon Sciences, 5232 Villigen PSI, Switzerland}

\author{Huibo Cao}
\affiliation{Neutron Scattering Division, Oak Ridge National Laboratory, Oak Ridge, Tennessee 37831, USA}

\author{Na Hyun Jo}
\email{nhjo@umich.edu}
\affiliation{Department of Physics, University of Michigan, Ann Arbor, MI 48109, USA}

\def\kill #1{\sout{#1}}
\def\add #1{\textcolor{blue}{#1}} 
\def\addred #1{\textcolor{red}{#1}} 

\date{\today}

\begin{abstract}
Altermagnets are a class of materials with compensated magnetic moments, in which spin sublattices are related by specific rotational symmetries other than inversion or translation. This allows time-reversal symmetry to be broken without a net magnetization. Cr-doped FeSb$_2$ has been theoretically proposed as a candidate \textit{d}-wave altermagnetic system, yet its magnetic ground state has remained unresolved. Here, we synthesize single crystals of Fe$_{1-x}$Cr$_x$Sb$_2$ and investigate their electrical transport and magnetic properties, with a focus on Fe$_{0.85}$Cr$_{0.15}$Sb$_2$. Magnetization measurements suggest spin-compensated ordering below $\sim$\,3.5~K, where magnetic moments align along the crystallographic \textit{b}-direction. Transport measurements reveal a crossover from large positive to negative magnetoresistance, while an anomalous Hall response emerges below 5~K, indicating time-reversal symmetry breaking. Muon spin relaxation measurements confirm that the magnetic ordering below 3.5~K is bulk in nature. The absence of coherent oscillations in zero-field $\mu$SR spectra and of magnetic Bragg intensity in single-crystal neutron diffraction establishes that the magnetically ordered state is short-range or disordered, rather than collinear altermagnetic order. These results demonstrate that Cr-doping alone breaks time-reversal symmetry without stabilizing long-range altermagnetic order in FeSb$_2$.

\end{abstract}

\maketitle

\section{Introduction}
Altermagnetism (AM) has recently emerged as a distinct class of collinear magnetism, extending beyond the conventional ferromagnetism (FM) and antiferromagnetism (AFM) \cite{Smejkal2022PRX,SmejkalPhysRevX_2}. These systems feature compensated magnetic moments, analogous to antiferromagnets, while global time-reversal symmetry is broken, as in ferromagnets. This behavior arises when spin sublattices with opposite, compensated moments are related by specific rotational (proper or improper and symmorphic or nonsymmorphic) operations, rather than simple translation or inversion \cite{SmejkalPhysRevX_2}. As a consequence, altermagnets can exhibit non-relativistic, spin-split electronic band structures along non-nodal planes with alternating spin polarization \cite{SpinSplitCollinearAFM,clustermultipole}. This symmetry-induced spin-splitting can yield nontrivial Berry curvature effects and can give rise to an anomalous Hall effect (AHE) \cite{TRsymmetry_SciAdv,MultipolarSpinSplit_PRB,_mejkal_2022}. These unique properties position altermagnets as promising candidates for spintronics and next-generation electronic devices \cite{AFMspintronic2018}.

Progress in discovery and understanding of AM systems has been driven largely by symmetry-based, theoretical analyses and first-principles predictions, which were later followed by experimental efforts \cite{RashbaZungerPRB2020,SmejkalPhysRevX_2,Chen_2025}. For example, the semiconducting altermagnet MnTe has been shown to exhibit spin-split electronic \cite{MnTe_ARPES,MnTE_ARPES_2,MnTE_ARPES_3} and magnonic bands \cite{MnTe_magnon_split,Chiral_MnTe_2}, as well as a strain-tunable AHE \cite{MnTe_thinfilm,Kluczykweakferro,smolenski2025,MnTe_gossamerferro,strain_MnTe_2}. The altermagnetic ground state has also been experimentally confirmed in other magnetic systems \cite{Fe2O3_magnon,Fe2O3_magnon_2}, including CrSb \cite{CrSb_NatCom,CrSb_PRL}, MnF$_2$ \cite{MnF2_neutrons}, Co$_{1/4}$NbSe$_2$ \cite{Regmi_2025,Co1/4NbS2_uSR}, and TbPt$_6$Al$_3$ \cite{TbPt6Al3_RE-ALM}. Despite recent advances, reconciling theoretical predictions with experimental realization in materials continues to pose significant challenges for altermagnetic systems \cite{RuO2_debate_review,KV2Se2O}. 

\begin{figure}
\centering
\includegraphics[width=0.7\linewidth]{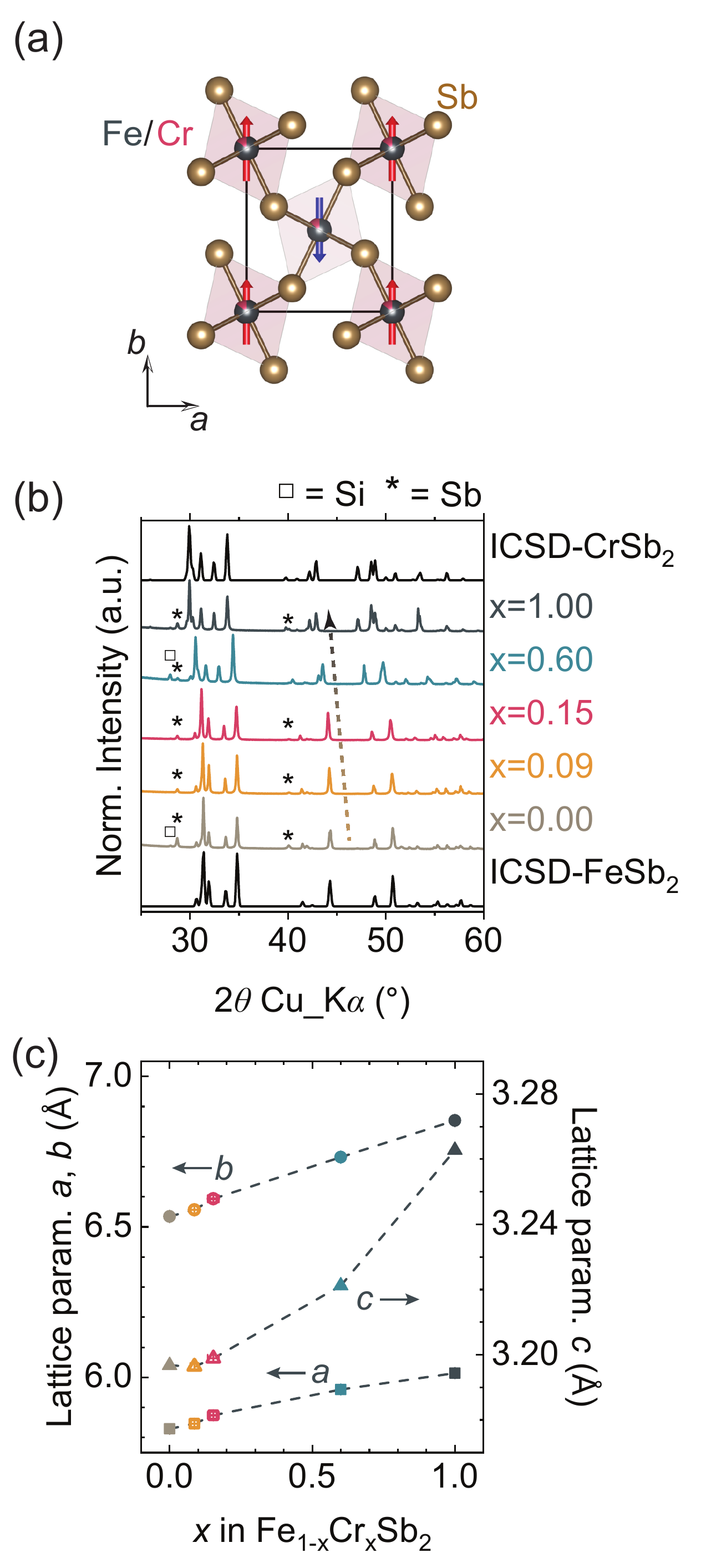}
\caption{(a) Crystal structure of orthorhombic Fe$_{1-x}$Cr$_x$Sb$_2$, viewed along the crystallographic $c$-direction, is overlaid with the proposed altermagnet magnetic structure where time-reversal symmetry is coupled to the rotational symmetry. (b) Powder X-ray diffraction patterns for each composition confirm the marcasite structure, with minority contributions from surface Sb-flux ($\star$) and Si sample holder ($\square$). With increasing Cr-content, peaks shift to lower $2\theta$ angles, confirming successful doping. (c) Lattice parameters, collected using Rietveld refinement, progressively increase with increasing Cr-content, further confirming homogenous doping. Dashed lines serve as a guide to the eye.}
\label{fig:structure}
\end{figure}

A fundamental distinction between AM and conventional AFM is that, in altermagnets, rotational/mirror symmetry (arising from spatial anisotropy in the crystal structure or electronic correlations) must be coupled with time-reversal symmetry at specific magnetic sites \cite{TRsymmetry_SciAdv,Mazin_PNAS2021}. The compounds Fe$_{1-x}$Cr$_x$Sb$_2$, which crystallize in the marcasite-structure type, serve as illustrative examples [see Fig. \ref{fig:structure} (a)]. The marcasite lattice hosts two distinct transition metal sites that are linked by a rotational symmetry operator. However, FeSb$_2$ has a weakly diamagnetic ground state \cite{FeSb2_localmoment_1,FeSb2_localmoment_2} and has been further described as a ``correlated $d$-electron topological Kondo-insulator'' \cite{doi:FeSb2_ARPES_PNAS2020,FeSb2_Kondo_PRB2005}. In contrast, CrSb$_2$ displays collinear antiferromagnetic ordering with $k = (1,0,0)$. However, the global coupling between rotational and time-reversal symmetries is not satisfied in CrSb$_2$ \cite{CrSb2_neutron}, resulting in conventional antiferromagnetism rather than altermagnetic order. Recently, Mazin \textit{et al.} proposed that substitution on the Fe site could stabilize compensated magnetic configurations compatible with altermagnetic symmetry [see Fig. \ref{fig:structure} (a)] \cite{Mazin_PNAS2021}, generating significant interest \cite{AHE_FeSb2_theory,FeSb2_torque,Co-dopedFeSb2_AMclaim}. In particular, Cr substitution has been suggested as a route to realizing such states, potentially accompanied by an anomalous Hall response. Prior experimental work has shown that Cr doping induces magnetic order in FeSb$_2$ \cite{Cr-doped_FeSb2}. However, despite these predictions, the magnetic ground state of Cr-doped FeSb$_2$ has remained unresolved. In particular, it remains unknown whether Cr substitution stabilizes long-range altermagnetic order in FeSb$_2$ in the manner required for AM, and whether it produces a measurable time-reversal symmetry breaking signature.

In this work, we address this problem by combining electrical transport, magnetization, muon spin relaxation ($\mu$SR), and neutron diffraction on Fe$_{1-x}$Cr$_{x}$Sb$_2$. Through these experiments of both macroscopic and local probes, we show that Fe$_{0.85}$Cr$_{0.15}$Sb$_2$ hosts bulk magnetic order below the magnetic ordering temperature, $T_N \sim$3.5~K. We also report the observation of an anomalous Hall effect, establishing broken time-reversal symmetry. However, using single-crystal neutron diffraction and $\mu$SR, we show that long-range magnetic ordering is absent in the Fe$_{0.85}$Cr$_{0.15}$Sb$_2$ samples. Our results narrow the magnetic ground state in Fe$_{0.85}$Cr$_{0.15}$Sb$_2$ system to short-range magnetic ordering, demonstrating that Cr doping alone does not stabilize the predicted long-range collinear altermagnetic state in the FeSb$_{2}$ system.  

\section{Methods}

\subsection*{Crystal growth and structural characterization}
Single crystals of Fe$_{1-x}$Cr$_x$Sb$_2$ were grown using a conventional solution method. High purity Fe (pieces, 99.98\%, Sigma Aldrich), Cr (pieces, 99.997\%, Sigma Aldrich), and Sb (shots, 99.9999\%, Thermo Scientific) were loaded into Canfield crucible sets \cite{CCS,lspceramics}, sealed in quartz tubes under 1/4 atmosphere of argon, and heated to 1000~$^{\circ}$C for 4 hours. The melt was cooled to 800~$^{\circ}$C over 10~h, followed by slow cooling to 650~$^{\circ}$C, where excess flux was decanted. A reduced cooling rate of 0.4~$^{\circ}$C\,h$^{-1}$ was necessary to obtain large crystals. Compositions were tuned to account for Fe-rich incorporation tendencies, with nominal ratios of 5.1:0.9:94 and 4.2:1.8:94 for $x=0.09$ and 0.15, respectively, and 2.4:3.6:94 for $x=0.60$. Final compositions were verified by energy-dispersive spectroscopy (SEM, TESCAN MIRA-3).

Crystal morphology depends on composition: Fe-rich samples ($x\leq0.09$) grow along [010] with [101] facets, while Cr-rich samples ($x\geq0.60$) favor [001] growth with [110] facets. Crystals with composition $x = 0.15$ displayed a mixture of both growth tendencies. Crystal orientation was determined using single-crystal X-ray diffraction (Rigaku XtaLAB Synergy, Mo K$\alpha$, $\lambda=0.71$~\AA). Data were collected via $\omega$ scans and processed with CrysAlisPro (version 1.171.42.101a, Rigaku OD, 2023), including Lorentz and polarization corrections. Powder X-ray diffraction (Rigaku MiniFlex, Cu K$\alpha$, $\lambda=1.54$~\AA) was performed on crushed crystals, and lattice parameters were obtained via Rietveld refinement using GSAS-II \cite{GSASII}. Structural visualization employed VESTA \cite{VESTA}.

\subsection*{Electrical and magnetic property measurement}

Electrical transport measurements were performed using a Quantum Design Physical Property Measurement System (PPMS). Crystals were shaped into rectangular bars ($\sim 700 \times 500 \times 200  ~\mu$m), and Pt wires were attached using H20E silver epoxy in standard four-probe and Hall configurations on [101] and [110] facets. For Fe$_{0.85}$Cr$_{0.15}$Sb$_2$ samples, sample surfaces were patterned with Al foil and gold-coated by sputtering before Au wires were attached using H20E silver epoxy. Resistivity ($\rho$) was measured with currents ($I$) of 0.5–3~mA along the [010] or [001] direction, under magnetic fields ($\mu_0H$) up to 14~T. Magnetoresistance and Hall measurements were conducted under field-cooled conditions, and the antisymmetric Hall component was extracted following standard procedures \cite{AHE_procedure}. Magnetization measurements were performed using the PPMS Vibrating Sample Magnetometer (VSM) and Magnetic Property Measurement System (MPMS). Crystals (1.5–45~mg) were mounted using GE varnish. Temperature-dependent susceptibility was measured from 2-300~K under field-cooled (FC) conditions (0.1–1~T), and field-dependent magnetization (\textit{M}) was recorded isothermally on FC samples. 

\subsection*{Muon spin relaxation}
Muon spin relaxation experiments were performed on the General Purpose Surface-Muon (GPS) instrument \cite{GPS_PSI} at the Swiss Muon Source (PSI). Single-crystal samples ($\sim 3 \times 2 \times 0.7$~mm) with aligned crystallographic axes were arranged in a mosaic configuration within a $7\times7$ mm grid. The muon spin polarization was rotated by 45$^{\circ}$ to ensure sensitivity to all directions. Measurements were conducted between 1.5–200~K in zero field, weak transverse field (3~mT), and longitudinal fields up to 0.2~T. Data were analyzed using the MUSRFIT package \cite{Musrfit}. 

\subsection*{Neutron diffraction}
Single-crystal neutron diffraction measurements were performed at the wide-angle single-crystal diffractometer (HB-2C WAND$^{2}$) at the High Flux Isotope Reactor (HFIR), Oak Ridge National Laboratory (ORNL). The wide 2D position sensitive detector arrays covers a Q-range of 0.37 – 7.50 Å$^{-1}$ (d-spacing: 17 – 0.84\AA) using unpolarized monochromatic neutrons with a constant wavelength of 1.486\AA. Reciprocal space maps were collected at temperatures of 1.5 K, 10 K, and 100 K, each with an angular coverage of 180° using a rotation step size of 0.1°, and a counting time of 10 s per step. The data were reduced and integrated using MantidWorkbench \cite{Mantid}. Order parameters of Bragg peaks (0 1 0) and (0 2 0) were measured over the temperature range of 1.5-100~K. Follow-up measurements were performed at the versatile intense triple-axis spectrometer (HB-1A VERITAS), HFIR, and at the elastic diffuse scattering spectrometer (BL-9 CORELLI) at the Spallation Neutron Source (SNS), ORNL.

\section{Results}
\subsection*{Crystal structure and sample morphology}
The marcasite crystal structure consists of corner-sharing Sb octahedra surrounding the transition metal site. The two metal-Sb octahedra are connected by a combination of rotation (180$^\circ$ around the \textit{b}-axis) and translation ($0,\frac{1}{2},\frac{1}{2}$). Powder X-ray diffraction (PXRD) patterns, collected on crushed crystals with varying $x$ in Fe$_{1-x}$Cr$_{x}$Sb$_2$, confirm single-phase marcasite structure at each composition [see Fig. \ref{fig:structure} (b)]. Only minor contributions were observed from elemental Sb (from flux; see \textit{Methods} for details on the growth procedure) and Si (from sample holder). With increasing Cr-content, the peak positions shift to lower $2\theta$ angles, consistent with the larger ionic radius of Cr compared to Fe. Lattice parameters were extracted using Rietveld refinement of the PXRD patterns, which are shown in Fig. \ref{fig:structure} (c). The parameters $a$ and $b$ follow Vegard's law, exhibiting a linear increase with increasing Cr-content, consistent with the observations of Hu \textit{et al.} \cite{Cr-doped_FeSb2}. However, the lattice parameter $c$ shows a non-monotonic trend; incorporation of small amounts of Cr leads to an initial decrease in $c$, followed by an increase at higher Cr concentrations. This behavior was observed in both single-crystal XRD and powder diffraction results. Overall, the increasing lattice parameters and shifting peak positions confirm the successful doping of Cr at the Fe-site. The crystal compositions were further confirmed using energy-dispersive spectroscopy [see Table S1]. In this report, we will focus on the electrical and magnetic properties of the Fe$_{0.85}$Cr$_{0.15}$Sb$_2$ samples, while the properties of all other Fe$_{1-x}$Cr$_{x}$Sb$_2$ samples are shown in the Supplemental Information file.

\begin{figure*}
\centering
\includegraphics[width=1\linewidth]{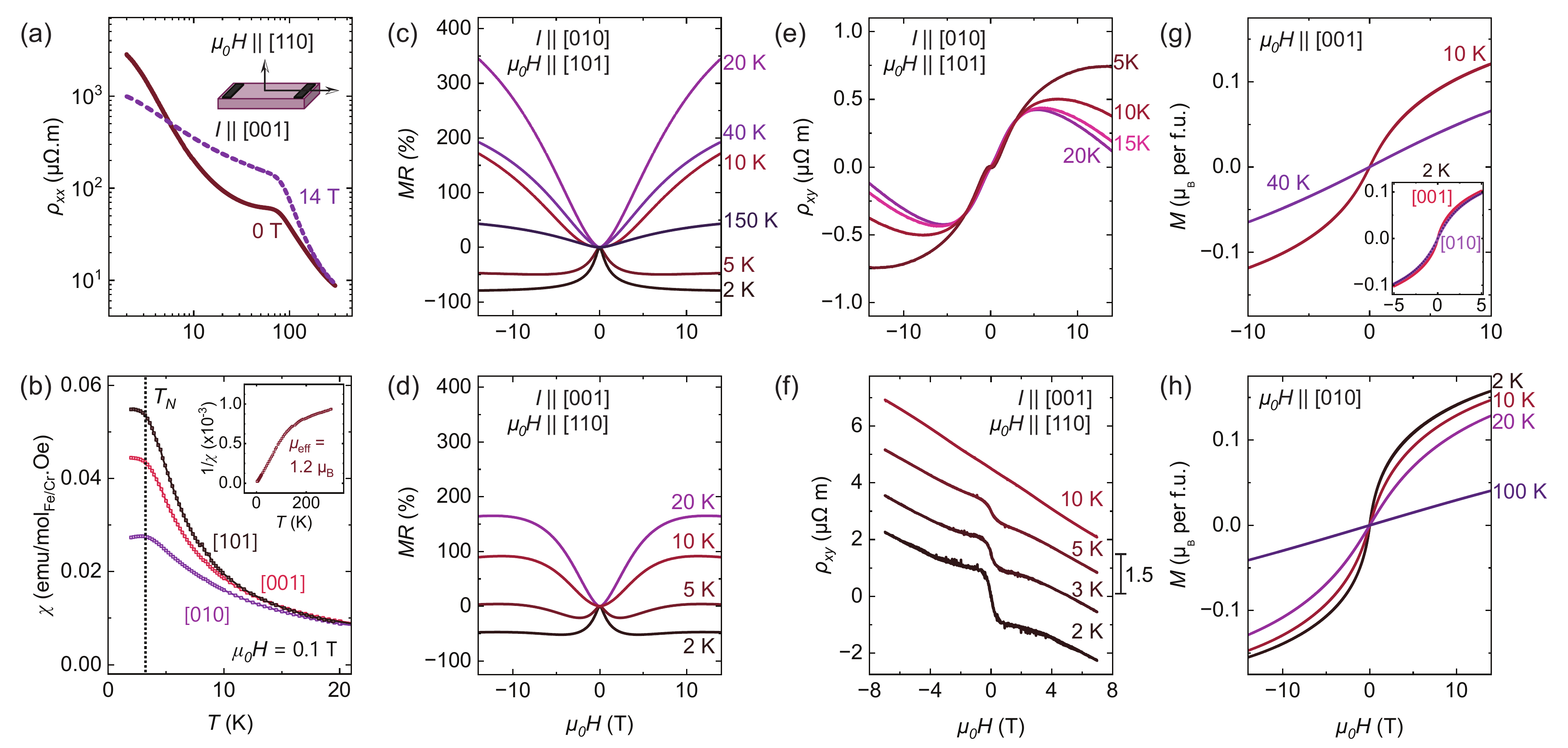}
\caption{Electrical and magnetic properties in the Fe$_{0.85}$Cr$_{0.15}$Sb$_2$ samples: (a) Temperature-dependent resistivity measured along [001] is shown under zero (solid lines) and under 14~T (dashed lines) perpendicular magnetic field applied along [110]. (b) Temperature-dependent magnetic susceptibility is shown for fields applied along [010], [001], and [101] measured with $\mu_0H$ = 0.1~T. A Curie plot is shown in the inset. (c) MR measured along [010] ($\mu_0H$ along [101]) and (d) MR measured along [001] ($\mu_0H$ along [110]) are shown, highlighting anisotropy. (e) Hall resistivity with \textit{I} along [010] ($\mu_0H$ along [101]) is shown measured on a sample with $t \simeq 160~\mu$m. (f) Hall resistivity with current along [001] (magnetic field along [110]) are shown at several temperatures separated by a constant offset of 1.5~$\mu\Omega~m$ measured on a sample with $t \simeq 500~\mu$m. (g) shows field-dependent magnetization measured along [001], while (h) shows $M(H)$ along [010]. $M(H)$ along both [010] and [001] at 2~K is shown in the inset of (g).}
\label{fig:transport}
\end{figure*}

\subsection*{Temperature-dependent electrical and magnetic properties}

The temperature dependence of the longitudinal resistivity $\rho_{xx}(T)$ for Fe$_{0.85}$Cr$_{0.15}$Sb$_2$ is shown in Fig. \ref{fig:transport} (a), where the current (\textit{I}) is applied along [001]. In the absence of the applied magnetic field ($\mu_0H$), the resistivity increases by more than two orders of magnitude as the temperature drops from 300~K to 2~K. Additionally, a broad cusp-like feature is observed at $\sim$100~K. Similar features are also observed in other Fe$_{1-x}$Cr$_{x}$Sb$_2$ samples [see Fig. S2]. An additional low-temperature cusp, which is visible in samples with $x = 0.0, 0.09,$ and $1.0$ below 10~K, was not clearly observed in the Fe$_{0.85}$Cr$_{0.15}$Sb$_2$ within the measured temperature regime. However, subtle changes in slope and a tendency toward saturation are evident at low $T$. These cusp-like features observed across the Fe$_{1-x}$Cr$_{x}$Sb$_2$ series are consistent with prior reports on FeSb$_2$-based systems and are typically attributed to thermally activated transport and surface states, respectively \cite{Takahashi_FeSb2_Hall,FeSb2_Kondo_PRB2005,FeSb2_localmoment_1, Sun2010_TE,Bentien_2007,Cr-doped_FeSb2}. Upon application of a perpendicular 14~T magnetic field along [110], a large positive magnetoresistance (MR) develops over a broad temperature range, and a linear $\rho_{xx}(T)$ dependence emerges below 80~K (note the log scale). On cooling below $\sim$ 5~K, MR shifts from positive to negative. This magnetic field dependence of $\rho_{xx}(T)$ is also observed when \textit{I} is applied along [010] ($\mu_0H$ applied along [101]) [See Fig. S2]. 

The magnetic susceptibility, $\chi(T)$, is shown in Fig. \ref{fig:transport} (b) measured with a 0.1~T magnetic field applied along the [001],[010], and [101] direction. $\chi(T)$ increases with decreasing temperature in a Curie-Weiss manner before reaching a maximum at $T_N$ = 3.5~K. The maximum $\chi$ along [001] is $\sim$ 60~\% greater than that along [010]. Furthermore, below $T_N$, $\chi(T)$ saturates along [001], while it decreases along [010]. These results suggest spin-compensation behavior (AFM-like) with the magnetic easy axis aligned along the crystallographic \textit{b-}direction. A Curie-Weiss fit ($1/\chi(T)$) below 150~K yields $\theta_{\mathrm{CW}}$ of $\sim$ 1~K and an effective magnetic moment of 1.2 $\mu_B$ per formula unit (f.u.). Note that a change in slope is also observed in the Curie plot (see inset of Fig. \ref{fig:transport} (b)) at $\sim$\,150~K, which likely stems from a low-spin-high-spin mixing of the $3d^6$ Fe-configuration previously discussed in the literature \cite{FeSb2_Kondo_PRB2005,Sun2011,FeSb2_spec_PNAS2024}. This feature is more apparent in the $x = 0.09$ samples [See Fig. S5].

\subsection*{Magnetic field-dependent physical properties}
MR along the two different crystallographic directions is shown in Fig. \ref{fig:transport} (c) and (d). A sizable positive MR is observed at 150~K, which gradually increases to reach a large, non-saturating MR ($\sim$340~\%) at 20~K along [010]. Upon further cooling, the MR reduces and turns negative between 10~K and 5~K. At 2~K, a sizable saturated negative MR of 80~\% is observed. Along [001], a maximum MR of 160~\% is observed at 20~K, which saturates at 10~T (see also Fig. S2 (b)). Overall, a lower MR magnitude was observed along [001] compared to that along the [010] direction, although negative MR is still observed between 10 and 5~K. 

Transverse component of Hall resistivity, $\rho_{xy}$, measured with \textit{I} along the [010] and [001] directions, with perpendicular $\mu_0H$ along [101] and [110], is shown in Fig. \ref{fig:transport} (e) and (f), respectively. The Hall voltage was measured perpendicular to both the current and magnetic field directions (see Fig. S4 for schematic). The symmetric and antisymmetric components were decoupled using a symmetrization procedure detailed in Ref. \cite{AHE_procedure}.  $\rho_{xy}$ exhibits complex, non-linear field-dependence in Fig. \ref{fig:transport} (e), which was measured on a sample with thickness (\textit{t}) of $\sim$ 160$~{\mu}$m. Field-induced non-linearity can be described using a two-band model, and has been previously observed in undoped FeSb$_2$ \cite{Takahashi_FeSb2_Hall}.  With \textit{I} along [001] and perpendicular $\mu_0H$ along [110], linear field-dependence of $\rho_{xy}$ is observed at 10~K measured on a sample with $t \simeq 500~\mu$m, as shown in Fig. \ref{fig:transport} (f). Beginning at 5~K and below, a measurable anomalous Hall component emerges in the $\rho_{xy}(H)$. The anomalous Hall component strengthens further with decreasing temperature, indicating time-reversal symmetry breaking. Notably, this AHE onset temperature (5~K) lies above the bulk ordering temperature $T_N = 3.5$~K defined from $\chi(T)$, and we do not observe any hysteresis at zero field. Below 5~K, the symmetrized $\rho_{xy}(H)$ signal proves noisy, likely due to the high resistance and large negative MR contributions. Attempts at further noise reduction using coarse binning/interpolation can lead to artifactual signals. Conversely, reducing sample thickness leads to increased non-linearity in $\rho_{xy}(H)$. This thickness dependence of complex, non-linear Hall response may be related to surface state contributions to electronic transport (see Section S2 for more details). This non-linear $\rho_{xy}(H)$, observed in Fe$_{0.85}$Cr$_{0.15}$Sb$_2$ samples, is less pronounced in the other Cr-doped FeSb$_2$ samples [see Fig. S3]. Furthermore, no AHE was observed in the other Cr-doped FeSb$_2$ samples.    
 
To further understand the magnetic ground state in Fe$_{0.85}$Cr$_{0.15}$Sb$_2$, field-dependent magnetization ($M(H)$) was measured with the magnetic field applied along [001] and [010], as shown in Fig. \ref{fig:transport} (g) and (h), respectively. Linear $M(H)$ is observed above 40~K, typical of paramagnetism at high temperature. Below 20~K, $M$ exhibits a sigmoidal field dependence with a positive slope along both directions. This shape persists below $T_N$ as well, as seen at 2~K in the inset of Fig. \ref{fig:transport} (g). At 2~K, a maximum $M$ of 0.16 $\mu_B$ per f.u. is recovered along [010] with 14~T field. The recovered $M(H)$ at 14~T corresponds to $\sim13\%$ of the $\mu_\textit{eff}$ calculated from the Curie fit, and $\sim 8\%$ of the $\mu_\textit{eff}$ value in the antiferromagnetic end member, CrSb$_2$ ($\mu_\textit{eff} = 1.94\, \mu_B$ per f.u.) \cite{CrSb2_neutron,CrSb2_Holseth1970,Cr-doped_FeSb2}. This small moment is consistent with a compensated magnetic ground state in Fe$_{0.85}$Cr$_{0.15}$Sb$_2$.

\subsection*{Local magnetic order using muon spin relaxation}
To further investigate the nature of the magnetic ordering, muon spin relaxation ($\mu$SR) measurements were performed on these samples. $\mu$SR has been successfully used to probe magnetic structures in other altermagnetic materials \cite{hicken2025MnTe,Co1/4NbS2_uSR}. In $\mu$SR, spin-polarized positive muons with a mean lifetime of $\sim$ 2.2~$\mu$s are implanted into the sample, and their decay positrons are counted over time. The implanted muon spin (spin-$\frac{1}{2}$) precesses at a frequency of $\omega_\mu = \gamma_\mu B_\text{internal}$, where $\gamma_\mu$ is the muon gyromagnetic ratio and $B_\text{internal}$ is the internal magnetic field of the sample. The decay positrons are preferentially emitted in the direction of the muon spin at the time of decay, allowing for a time-dependent polarization measurement of the muon. Since the implanted muons are sensitive to the local magnetic field around them, this measurement serves as a probe for local magnetic field. In the weak transverse field (TF) configuration, a small magnetic field ($B_\text{external}$) is applied perpendicular to the muon polarization, causing the muon ensemble to precess symmetrically in the absence of local magnetic order ($T>T_c$). If magnetic order is present ($T \leq T_c$), a net $B_\text{internal}$ leads to a loss of precessional symmetry in the TF signal when intentionally measured with a low time resolution.  

\begin{figure*}
\centering
\includegraphics[width=1\linewidth]{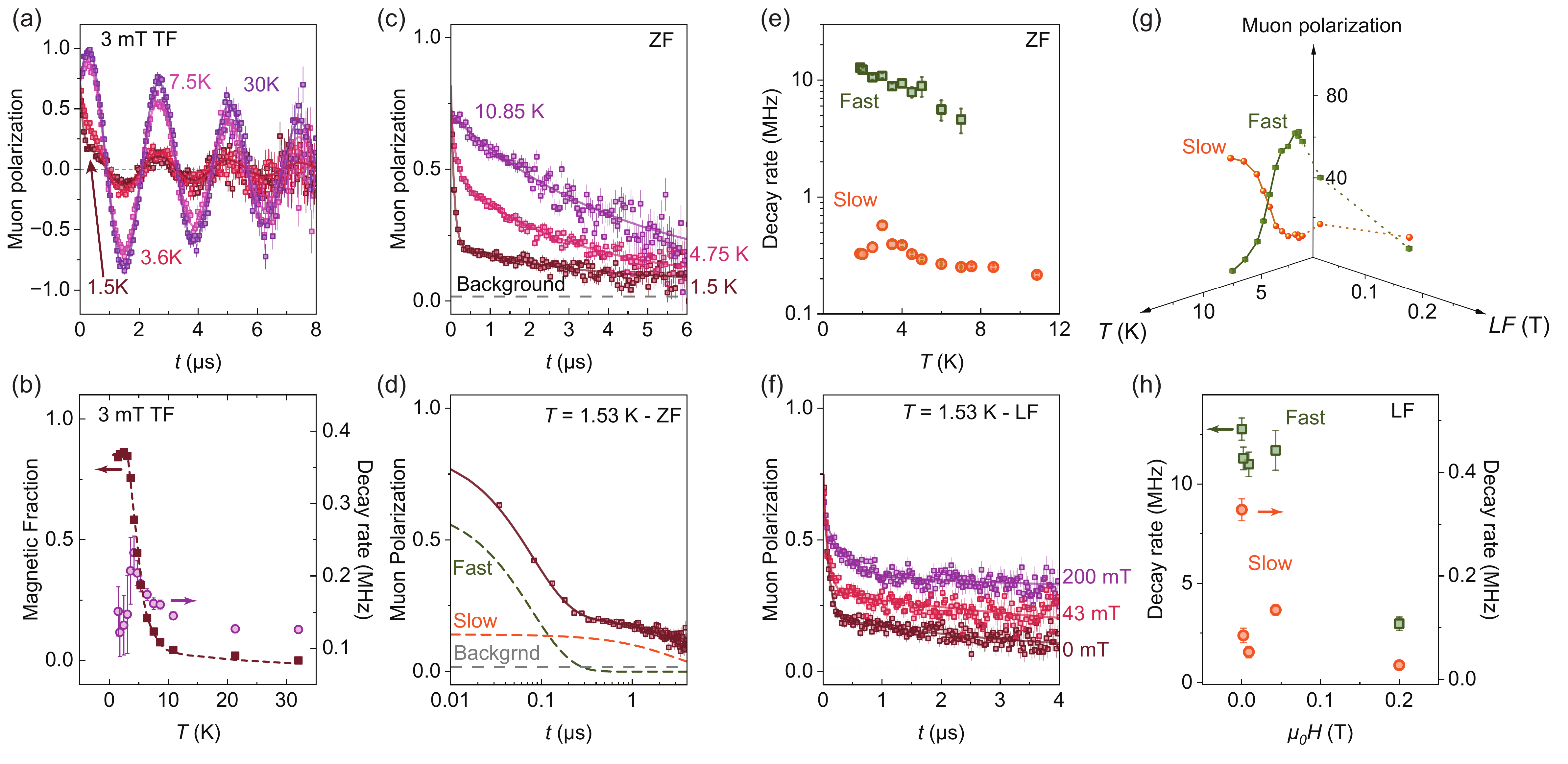}
\caption{(a) Weak transverse field (TF) muon spectra reveal asymmetry at 7.5~K and below. (b) Magnetic volume fraction, extracted by fitting weak transverse field spectra, shows nearly the entire sample is magnetically ordered below $T_N$. (c) Zero field (ZF) $\mu$SR spectra also reveal muon polarization loss without oscillations. (d) ZF spectrum at 1.53~K shows two distinct decay rates, one fast and one slow, whose temperature evolution is shown in (e). (f) An applied longitudinal magnetic field up to 0.2~T decouples the fast component, causing muon polarization to increase with field. (g) Temperature and magnetic field dependence of muon polarization for the two components are shown, and their field-dependent decay rates are shown in (h).}
\label{fig:muSR}
\end{figure*}

Fig. \ref{fig:muSR} (a) shows representative time-dependent $\mu$SR spectra collected under a weak TF of 3~mT on co-aligned Fe$_{0.85}$Cr$_{0.15}$Sb$_2$ single crystal samples. At 30~K, the samples exhibit complete muon polarization, suggesting the absence of any magnetic ordering, consistent with paramagnetism. At 7.5~K, a partial polarization loss is observed, and by 3.6~K, a significant polarization loss is evident. The time-domain polarization ($P(t)$) was fitted using the equation,

\begin{equation}
    P_\text{TF}(t) =\frac{A(t)}{A_\text{max}}  = P_\text{para}\cos(2\pi\nu t + \phi) e^{-\lambda_\text{TF}t }
    \label{TF_fit}
\end{equation}

where the muon polarization, $P_\text{TF}(t)$, is the normalized detector asymmetry, $A(t)/A_\text{max}$, $P_\text{para}$ is the polarization of the paramagnetic component, $\nu$ is the muon precession frequency, $\phi$ is the phase factor, and $\lambda_\text{TF}$ is the muon spin relaxation rate. Suppression of the $P_\text{para}$ upon cooling indicates magnetic ordering, and the ordered magnetic fraction is given by $1-P_\text{para}$, as shown in Fig. \ref{fig:muSR} (b). The results indicate that magnetic order develops over a broad temperature range below 10~K, with the whole sample ordering by 3.6~K. The broad temperature range is consistent with the gradual ordering of the randomly dispersed Cr dopants \cite{Cr-doped_FeSb2}. We define $T_N = 3.5$~K as the temperature at which bulk susceptibility saturates and the full sample volume orders. The temperature-dependent muon decay rate exhibits a peak at $T_N$, which is consistent with a phase transition. Note that we further confirmed paramagnetic behavior in $x = 0.09$ samples and antiferromagnetic ordering in $x = 0.60$ samples using a combination of $\mu$SR and magnetization measurements, which are shown in Fig. S5.  

In an attempt to elucidate the magnetic structure, we further performed $\mu$SR measurements in the absence of an external magnetic field (ZF configuration); representative spectra are shown in Fig. \ref{fig:muSR} (c). No oscillations were detected in the magnetically ordered state down to 1.53~K. Furthermore, the muon polarization decays quickly, with two decay components: one fast and one slow. The muon polarization spectra ($P_\text{ZF}(t) = A_\text{ZF}(t)/A_\text{max}$) were therefore fit using the equation,
\begin{equation}
    P_\text{ZF}(t) = P_{\text{Fast}}e^{-\lambda_{\text{Fast}}t } +P_{\text{Slow}}e^{-\lambda_{\text{Slow}}t} +P_{\text{Background}}
    \label{ZF_fit}
\end{equation}
where $P_{\text{Fast}}$ and $P_{\text{Slow}}$ represent the fractions of each component, while $\lambda_{\text{Fast}}$ and $\lambda_{\text{Slow}}$ are the respective decay rates. A constant background $P_\text{Background}$ of 7~\% was used at all temperatures to account for muons stopping outside the sample, which is within the typical background detected at the General Purpose Surface-Muon Instrument (GPS) \cite{GPS_PSI,hicken2025MnTe}. Fig. \ref{fig:muSR} (d) shows the fit for spectra collected at 1.53~K. At this temperature, the fast component comprises $\sim$ 68~\% (i.e., 0.1555/0.229) of the muon polarization in the sample, with the slow component making up the rest. Muon polarization is typically split into 1:2 ratio of fast and slow components in powder samples with randomly oriented domains. Reaching the powder limit in these single crystal samples suggests that the magnetic ordering present is relatively short-range. As temperature increases, the fast component is suppressed while the slow component dominates, as seen in Fig. \ref{fig:muSR} (g). The temperature-dependence of the two decay rates is shown in Fig. \ref{fig:muSR} (e). The slow component decay rate exhibits a peak at $T_N$, which is characteristic of a dynamic component \cite{2Component,2_component_2,2_component_3}. Conversely, the fast component decay rate, which is an order-of-magnitude faster, decreases with increasing temperature and behaves like an order parameter for (quasi-)static magnetism \cite{2Component,2_component_2,2_component_3}. Above $T_N$, the fast component contributions become small and difficult to separate, as seen in Fig. \ref{fig:muSR} (g). The lack of oscillations in ZF spectra below $T_N$ further suggests a relatively \textit{disordered} magnetic ground state, which further support a short-range magnetic ordering picture in Fe$_{0.85}$Cr$_{0.15}$Sb$_2$. A non-uniform distribution of the internal magnetic fields can cause an absence of oscillations in ZF, which is expected in a doped sample with randomly dispersed Fe/Cr moments. Instead, a quickly decaying static component emerges, as the $B_\text{internal}$ is smaller than the field distribution width ($\lambda_\text{Fast}/\gamma_\mu$) of $0.094 \pm 0.004$~T \cite{distr_width}. In theory, muon stopping site calculations could be done to estimate the magnetic correlation lengths, which may reveal the lower bound of Cr-doping required to trigger long-range magnetic ordering.  

Magnetic field-dependence of the $\mu$SR spectra was evaluated using longitudinal fields (LF) at the base temperature; representative spectra are shown in Fig. \ref{fig:muSR} (f). Under a longitudinal field, $P_\text{Fast}$ is suppressed dramatically, as shown in Fig. \ref{fig:muSR} (g). In contrast, the slow component strengthens and cannot be decoupled using LF. This field-dependence of the two components reiterates the expected quasi-static (dynamic) behavior of the fast (slow) components. The field-dependent decay rates are shown in Fig. \ref{fig:muSR} (h). Our $\mu$SR results confirm that the magnetic ordering in Fe$_{0.85}$Cr$_{0.15}$Sb$_2$ samples is bulk and short-range, and that ordering occurs gradually below $\sim$ 10~K. At $\sim$ 5~K, about half the sample exhibits magnetic ordering, which may lead to the observed time-reversal symmetry breaking in Fig. \ref{fig:transport} (f).

\begin{figure*}
\centering
\includegraphics[width=1\linewidth]{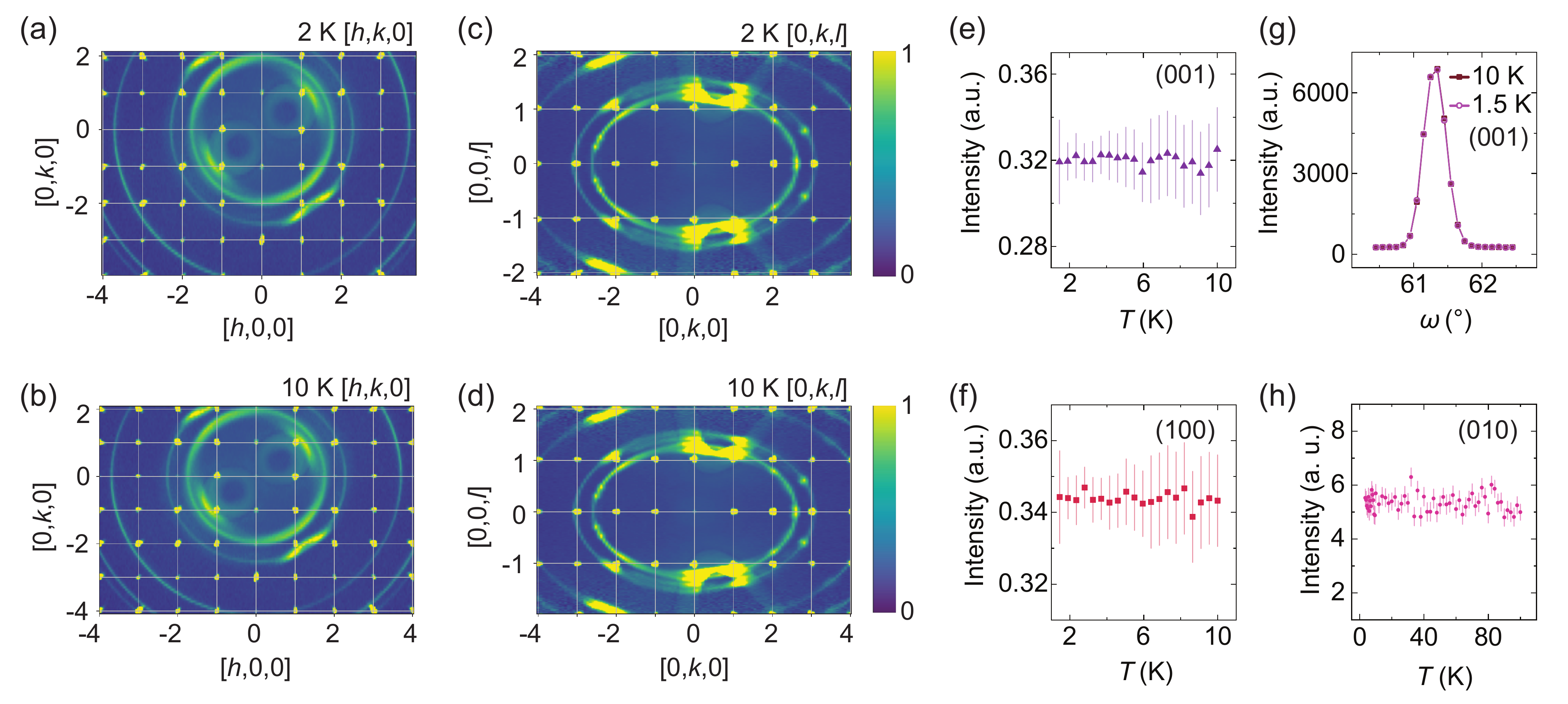}
\caption{Reciprocal space maps collected at 2~K ((a) and (c)) and 10~K ((b) and (d)) show that no additional peaks emerge below $T_N$. The intensities of symmetry-forbidden peaks, shown at (e) (001) and (f) (100), also show no temperature dependence. (g) Rocking curve scans of the (001) peak at 1.5~K and 10~K measured with a high-flux triple-axis instrument exhibit no intensity differences. (h) The temperature dependence of the (010) peak intensity rules out magnetic ordering at higher temperatures.}
\label{fig:neutron}
\end{figure*}

\subsection*{Neutron diffraction for magnetic structure elucidation}
Neutrons, due to their inherent magnetic moment, are sensitive to magnetic ordering. As a result, single-crystal neutron diffraction serves as a key tool to elucidate the underlying magnetic structure in samples \cite{neutron_diffraction}. Reciprocal space maps collected below and above the magnetic transition (Fig. \ref{fig:neutron}(a)-(d)) show no additional magnetic Bragg peaks at low temperature, establishing a magnetic propagation vector of $k = 0$ which is associated with the magnetic unit cell coinciding with the crystallographic unit cell. 

In $k = 0$ systems, the absence of new Bragg reflections prevents direct identification of the magnetic structure through conventional peak indexing. Lack of magnetic peaks below $T_N$ could emerge from a non-magnetic ground state, a FM ground state, or a $k = 0$ AM state. We attempted to fit the temperature-dependent changes in the integrated intensities of nuclear Bragg reflections, but no satisfactory fits could be achieved from the integrated intensities of 49 peaks. Alternatively, magnetic structure can be elucidated by investigating the magnetic order parameter along different crystallographic direction. In the Fe$_{0.85}$Cr$_{0.15}$Sb$_2$ samples, weak intensities are observed at certain symmetry-forbidden locations, including (010) and (100). Forbidden substructure peaks have been previously observed in FeSb$_2$ in both neutron and synchrotron X-ray diffraction \cite{vacancy-ordering}, and are attributed to local ordering of Sb vacancies that induce weak monoclinic distortions. In search of magnetic order parameter, we measured temperature-dependent integrated intensities of three symmetry-forbidden reflections where the nuclear structure factor contributions are expected to be the smallest, and where magnetic scattering would be most detectable within the single-crystal diffraction limit. No temperature-dependent intensity was observed at any of these reflections (Fig. \ref{fig:neutron} (e) and (f)), and rocking curve scans of the (001) peak at 1.5~K and 10~K on a high-flux triple-axis instrument confirm the absence of magnetic intensity across $T_N$ (Fig. \ref{fig:neutron} (g)). Furthermore, the temperature-dependence of the (010) peak intensity over a broader temperature range rules out any undiscovered magnetic ordering above $T_N$ (Fig. \ref{fig:neutron} (h)). The absence of magnetic intensity below $T_N$ suggests that long-range magnetic order is absent in the Fe$_{0.85}$Cr$_{0.15}$Sb$_2$ samples. However, local magnetic ordering (short-range) could not be ruled out through our measurements, considering its much lower scattering signal in a broader reciprocal space. Diffuse scattering on powder or co-aligned single crystal samples might be able to resolve any short-range magnetic correlations.

\section{Discussion and Conclusion}
Our combined measurements establish that Fe$_{0.85}$Cr$_{0.15}$Sb$_2$ hosts a short-range or disordered, spin-compensated magnetic ground state with broken time-reversal symmetry. $\chi(T)$ exhibits saturation along [001] and a cusp-like feature along [010], indicating that [010] is the magnetic easy axis in a spin-compensated system. Magnetization measurements demonstrate that the ordered moment is small and non-saturating up to 14~T, consistent with a compensated spin structure. $\mu$SR results under a weak TF establish that magnetic ordering occurs throughout the entire sample below $T_N$, ruling out surface-driven magnetism. Gradual ordering of randomly dispersed Fe/Cr atoms begins below 10~K, which is consistent with dilute, disordered magnetic dopants ordering incrementally; the sample reaches full magnetic volume fraction at $T_N = 3.5$~K, matching the bulk transition seen in $\chi(T)$. This gradual ordering also triggers an anomalous Hall signal emergence near 5~K (above $T_N$). Since roughly half the sample volume is already magnetically ordered at this temperature, it is sufficient to locally break time-reversal symmetry before the full sample orders. 

On the other hand, coherent oscillations are absent in the ZF-$\mu$SR spectra, suggesting that the magnetic correlations may be short-range or disordered. Instead, two decay components emerge reaching the powder limit in single crystal samples. Furthermore, magnetic intensities could not be resolved in single-crystal neutron diffraction along the three crystallographic directions. This further established the magnetically ordered state observed in magnetization and $\mu$SR as lacking long-range ordering expected from a collinear AM \cite{Mazin_PNAS2021}. 

In parallel, the observation of an anomalous Hall component in $\rho_{xy}(H)$ demonstrates that time-reversal symmetry is broken in the electronic structure. The two magnetic sublattices likely occupy crystallographic sites related by a 180$^\circ$ rotation about the \textit{b}-axis. This symmetry operation does not restore time-reversal symmetry, giving rise to non-zero Berry curvature \cite{Mazin_PNAS2021}. However, a collinear altermagnetic model does not completely describe the magnetic ground state. Instead, a short-range spin-compensated magnetic state with broken time-reversal symmetry is stabilized using Cr-doping alone. Future work varying the Fe/Cr ratio could further improve magnetic correlation length through tuning the dopant concentration, and potentially stabilize long-range altermagnetic order that may be useful for practical applications. More broadly, these results show that first-principles calculations, while promising, benefit from rigorous experimental validation using both bulk and local probes to further our understanding of altermagnetism.

\begin{acknowledgments}
The authors thank Prof. Kai Sun and Dr. Alon Avidor for insightful discussions. This work was supported from the National Science Foundation (NSF) through a CAREER grant (Award No. DMR-2337535). S.S. was supported by the Materials Research Science and Engineering Center at the University of Michigan (Award No. DMR-2309029). M.X. and W.X. at Michigan State University were supported by the U.S. Department of Energy (DOE), Division of Basic Energy Sciences (Award No. DE-SC0023648). E.G. gratefully acknowledges the funding through the Deutsche Forschungsgemeinschaft (DFG, German Research Foundation) through Grant No. TRR 288—422213477 (Project No A13). The magnetization measurements at the University of Michigan are supported by the Department of Energy under Award No. DE-SC0020184  to L.L. and Y. Z. The authors acknowledge the University of Michigan College of Engineering for financial support and the Michigan Center for Materials Characterization for use of the instruments and staff assistance. Part of this work is based on experiments performed at the Swiss Muon Source (S$\mu$S), Paul Scherrer Institute, Villigen, Switzerland. A portion of this research used resources at the High Flux Isotope Reactor and the Spallation Neutron Source, DOE Office of Science User Facilities operated by the Oak Ridge National Laboratory. The beam time was allocated to HB-2C WAND$^{2}$ on proposal number IPTS-37619, to CORELLI on proposal number IPTS-38396, and to HB-1A VERITAS on proposal number IPTS-38517.
\end{acknowledgments}

\end{document}


\maketitle
\newpage

\section{Morphology and composition}

\begin{figure*}[ht!]
\centering
\includegraphics[width=0.8\linewidth]{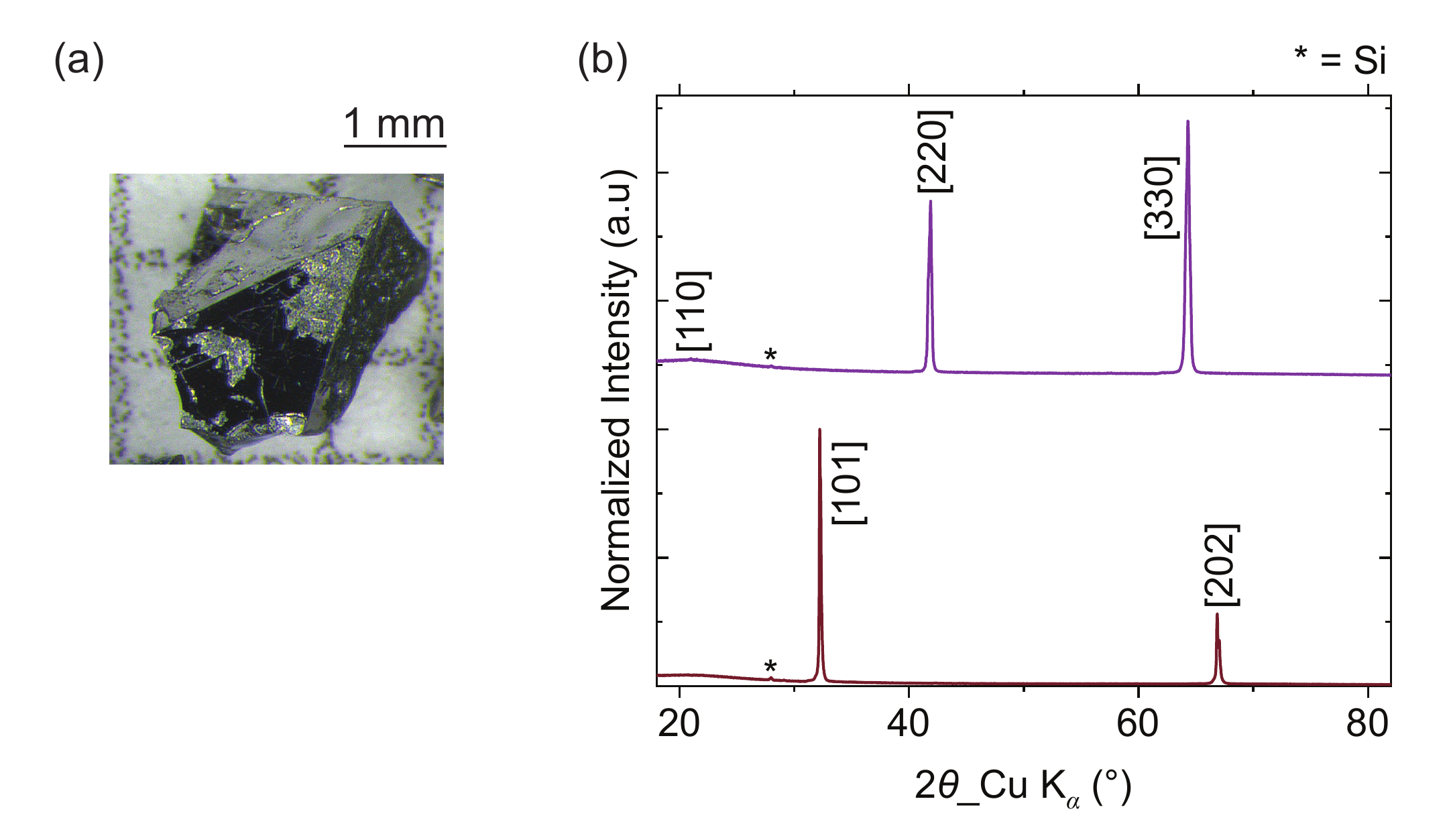}
\caption{(a) A representative Fe$_{0.85}$Cr$_{0.15}$Sb$_2$ single crystal is shown on a mm-grid paper. (b) X-ray diffraction pattern collected on single crystal sample shows peaks belonging to the [101] and [110] families. $\star$ represents Si diffraction peak coming from the sample holder.}
\label{fig:SI Crystal structure}
\end{figure*}

Single crystals of Fe$_{1-x}$Cr$_x$Sb$_2$ were grown using a conventional solution method using Sb-flux. Crystal compositions were determined using energy-dispersive spectroscopy (EDS) equipped on a scanning electron microscope (SEM). The nominal and final EDS compositions are shown in Table \ref{tab:EDS}. A representative single crystal with the composition Fe$_{0.85}$Cr$_{0.15}$Sb$_2$ is shown in Figure \ref{fig:SI Crystal structure} (a). These crystals exhibit well-formed [110] and [101] facets, which were identified through X-ray diffraction (XRD). Representative XRD patterns showing the [110] and [101] facets are shown in Fig. \ref{fig:SI Crystal structure} (b).

Fe$_{1-x}$Cr$_x$Sb$_2$ crystals grow in two distinct morphologies. Samples with Fe-rich compositions $x = 0.00$ and $0.09$ grew primarily along the [010] direction, exhibiting well-defined [101] facets. Cr-rich crystals, with $x = 0.60$ and $1.00$, grew predominantly along the [001] direction, where large [110] facets could be easily separated. Crystals with a composition $x = 0.15$ displayed a mixture of both growth tendencies.

\begin{table*}[ht!]
    \centering
    \begin{tabular}{c|c|c}
         $x$ in Fe$_{1-x}$Cr$_x$Sb$_2$ & Nominal composition & EDS composition \\
         0.00 & Fe$_{0.06}$Sb$_{0.94}$ & Fe$_{1}$Sb$_{2}$\\
         0.09 & Fe$_{0.051}$Cr$_{0.009}$Sb$_{0.94}$ & Fe$_{0.913\pm0.013}$Cr$_{0.087\pm0.013}$Sb$_{2}$\\
         0.15 & Fe$_{0.042}$Cr$_{0.018}$Sb$_{0.94}$ & Fe$_{0.847\pm0.017}$Cr$_{0.153\pm0.017}$Sb$_{2}$\\
         0.60 & Fe$_{0.024}$Cr$_{0.036}$Sb$_{0.94}$ & Fe$_{0.60\pm0.01}$Cr$_{0.40\pm0.01}$Sb$_{2}$\\
         1.00 & Cr$_{0.06}$Sb$_{0.94}$ & Cr$_{1}$Sb$_{2}$
    \end{tabular}
    \caption{Nominal and EDS compositions of Fe$_{1-x}$Cr$_x$Sb$_2$ crystals}
    \label{tab:EDS}
\end{table*}

\pagebreak

\section{Electrical transport properties of Fe$_{1-x}$Cr$_x$Sb$_2$ samples}

\begin{figure*}[ht!]
\centering
\includegraphics[width=0.5\linewidth]{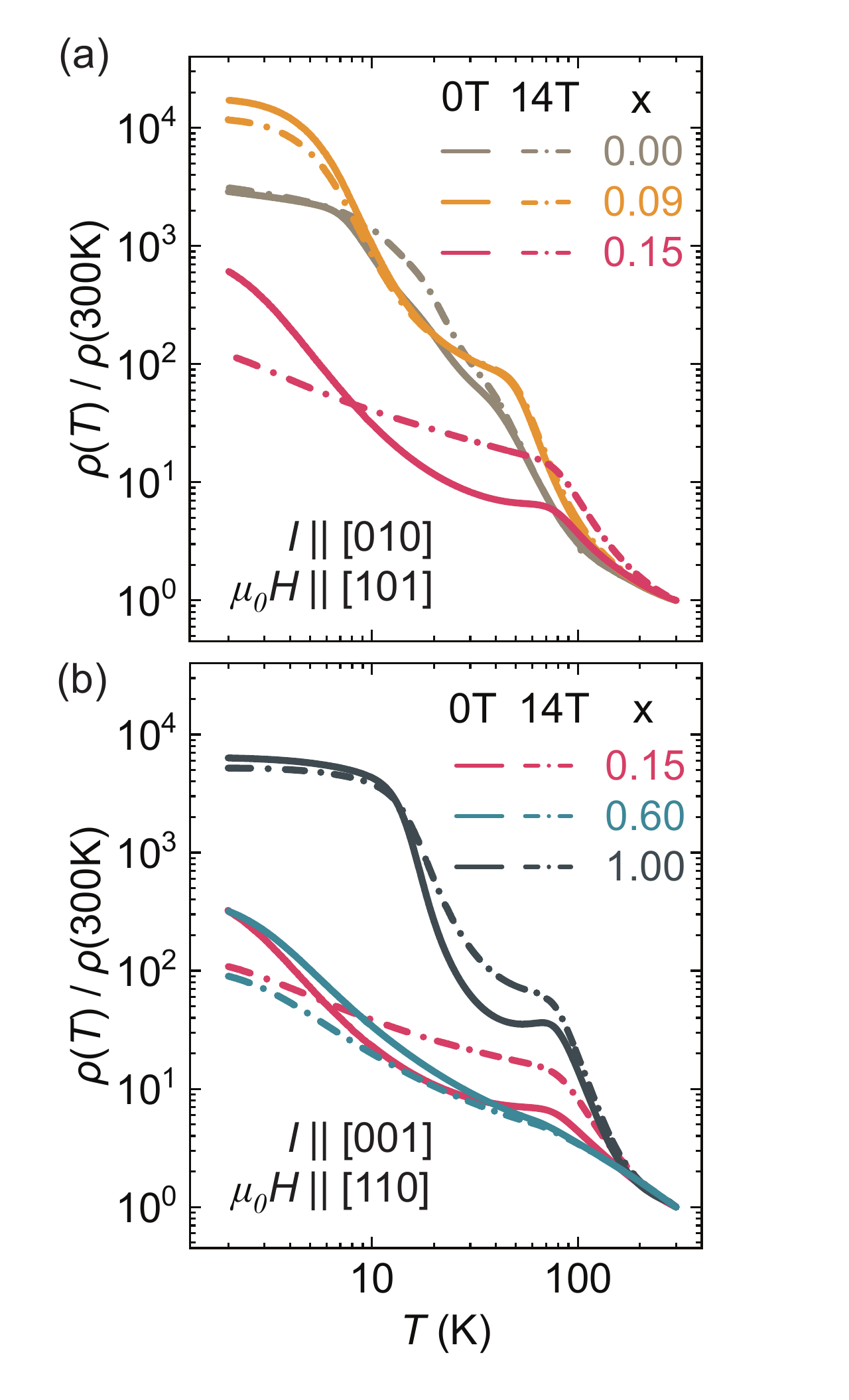}
\caption{(a) Resistivity measured along [010] is shown as a function of temperature under zero (solid lines) and 14~T (dashed lines) magnetic fields for samples with $x = 0.00, 0.09,$ and $0.15$ in Fe$_{1-x}$Cr$_x$Sb$_2$. The magnetic field was applied along the [101] direction. (b) Resistivity measured along [001] is shown for samples with $x = 0.15, 0.60,$ and $1.00$ under zero (solid lines) and 14~T (dashed lines) magnetic fields applied along the [110] direction.}
\label{SI Resistivity}
\end{figure*}

Electrical transport properties were measured with current and voltage contacts placed on the well-identified [101] and [110] facets. Fig. \ref{SI Resistivity} (a) shows the temperature-dependent $\rho_{xx}$ (normalized to resistivity at 300~K) where \textit{I} was applied along [010] under zero applied field (solid lines) and 14~T magnetic field (dashed lines) along [101] for $x = 0.00, 0.09,$ and $0.15$ samples. Normalized $\rho_{xx}(T)$ was measured with \textit{I} along [001] and $\mu_0H$ along [110] in samples with $x = 0.15, 0.60,$ and $1.00$, as shown in Fig. \ref{SI Resistivity} (b). With decreasing temperature from 300-2~K, $\rho_{xx}(T)$ increases by two to four orders of magnitude. Two distinct cusps are observed in samples with $x = 0.00, 0.09,$ and $1.00$, consistent with literature reports \cite{Takahashi_FeSb2_Hall,FeSb2_Kondo_PRB2005,FeSb2_localmoment_1,Sun2010_TE}. Only the high temperature cusp at $\sim$~100~K is visible in samples with $x = 0.15$. Both cusps are heavily suppressed in the heavily doped sample with $x = 0.60$.  

\begin{figure*}[h!]
\centering
\includegraphics[width=0.8\linewidth]{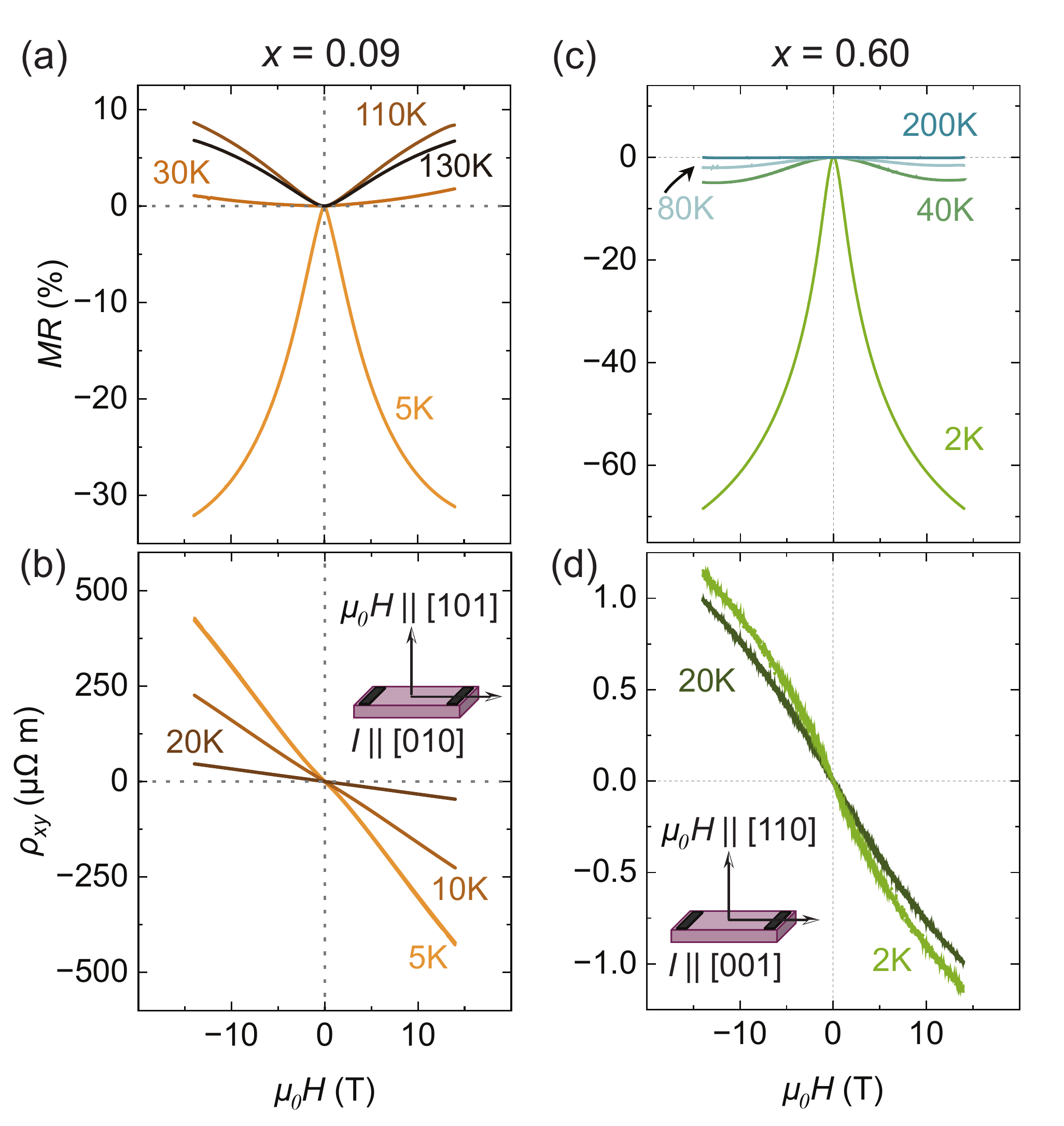}
\caption{Resistivity was measured as a function of applied magnetic field in both the longitudinal and Hall configurations. Field-dependence of (a) MR and (b) Hall resistivity for samples with $x = 0.09$ is shown. In $x = 0.60$ samples, (c) MR and (d) Hall resistivity were measured with current along [001] and magnetic field along [110].}
\label{SI MR+Hall}
\end{figure*}

MR and Hall resistivity were also measured in the doped samples. Hall voltage was measured perpendicular to both current and magnetic field directions. The antisymmetric component of Hall resistivity was extracted using the procedure outlined in Ref \cite{AHE_procedure}. Crystals with $x = 0.09$ show positive, non-saturated MR at 130~K, which increases modestly at 110~K, as seen in Fig. \ref{SI MR+Hall} (a). Upon further cooling, MR drops close to zero at 30~K and turns negative at even lower temperatures. A sizable, non-saturated negative MR of $\sim 33~\%$ is observed at 5~K under a 14~T magnetic field. In the Hall configuration, the antisymmetric component shows a linear field-dependence, as shown in Fig. \ref{SI MR+Hall} (b). The negative slope decreases with increasing temperature, suggesting thermal excitation of \textit{n}-type carriers, consistent with the $\rho_{xx}(T)$ behavior. 

In the Cr-rich samples with composition $x = 0.60$, negative MR appears below 90~K, with the magnitude of negative MR increasing linearly with decreasing temperature. As shown in Fig. \ref{SI MR+Hall} (g), a large negative MR of $\sim 70 \%$ is observed at 2 K, which remains non-saturating up to 14~T. Fig. \ref{SI MR+Hall} (h) shows the field-dependent Hall resistivity, which displays a negative slope, indicating that n-type carriers are dominant. The slope becomes steeper with decreasing temperature, suggesting a lower concentration of charge carriers at 2~K than at 20~K. A slight field-dependent curvature is observed in addition to a linear ordinary Hall effect, which could be due to could be associated with weak ferromagnetism due to spin canting \cite{Cr-doped_FeSb2} or multiband/surface effects.

\begin{figure*}
\centering
\includegraphics[width=0.5\linewidth]{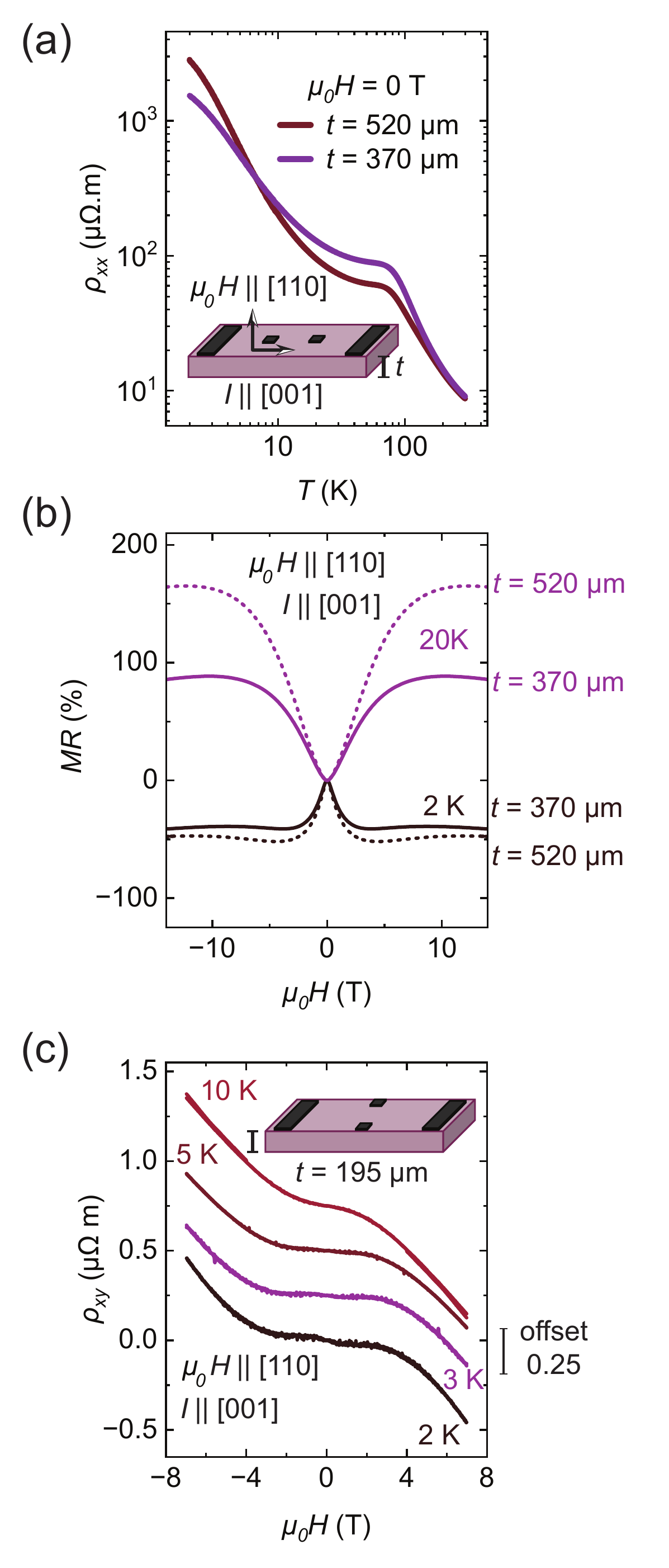}
\caption{(a) Resistivity was measured in the four-probe configuration with \textit{I} along [001] ($\mu_0H$ along [110]) on the same crystal at two different thickness. (b) MR measured at 20~K and 2~K show reduced MR with lower thickness. (c) In the Hall geometry, transverse resistivity measured on a crystal with $t = 195~\mu$m exhibits non-linear field-dependence at 10~K and below. A constant offset of 0.25~$\mu\Omega~m$ was used to separate different temperature points.}
\label{fig:surface_transport}
\end{figure*}

Contribution of surface states on the electrical transport properties was investigated by varying the sample thickness (\textit{t}). Samples with lower thickness have greater surface area-to-volume ratio, and therefore higher contribution of surface states on the electronic transport. This has been thoroughly investigated in undoped FeSb$_2$ \cite{doi:FeSb2_ARPES_PNAS2020,eaton2024electricaltransportsignaturesmetallic}. In the Cr-doped sample Fe$_{0.85}$Cr$_{0.15}$Sb$_2$, we measured four-probe resistivity and MR on the same sample at two different thickness, as shown in Fig. \ref{fig:surface_transport} (a) and (b) respectively. Here, \textit{I} was applied along [001] direction and a perpendicular $\mu_0H$ was applied along [110]. At lower \textit{t} (higher surface contributions), the sample exhibits higher resistivity between 200~K and 10~K, and lower resistivity below 8~K. On the application of a perpendicular magnetic field, lower magnitude of MR is observed with lower thickness. In the Hall configuration, shown in Fig. \ref{fig:surface_transport} (c), a sample with $t = 195~\mu$m exhibits pronounced non-linearity in $\rho_{xy}(H)$ when compared to that in a sample with $t \simeq 500~\mu$m in Fig. 2 (f) (main text).

\clearpage

\section{Magnetic properties in Cr-doped FeSb$_2$}

\begin{figure*}[h!]
\centering
\includegraphics[width=1\linewidth]{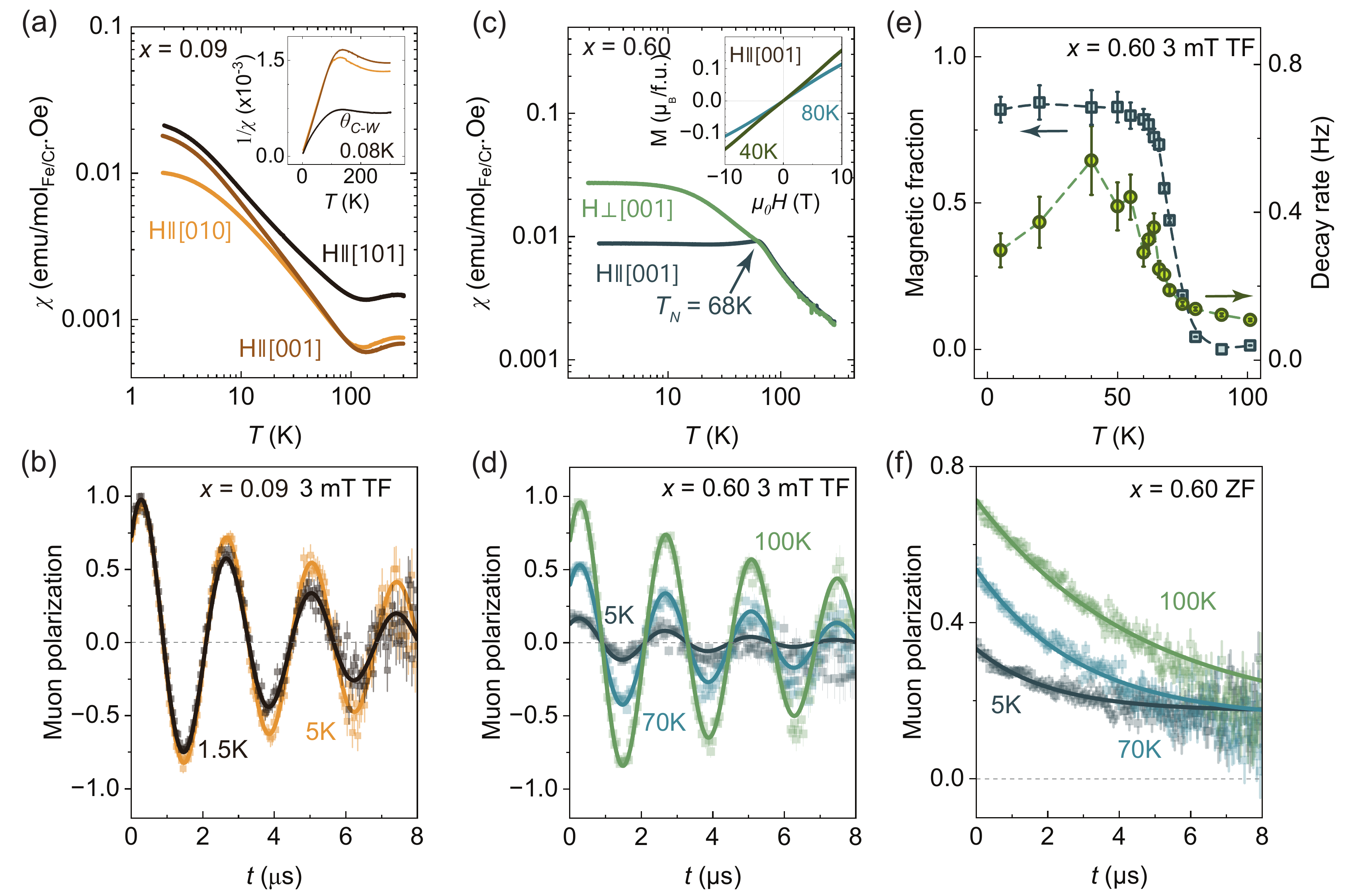}
\caption{(a) In samples with $x = 0.09$, magnetic susceptibility is shown as a function of temperature, where a 1~T magnetic field was applied along the [010], [001], and [101] directions. 1/$\chi$ vs T is shown in the inset. (b) Muon polarization measured under a weak 3~mT transverse field shows cosine time-dependence at base temperatures of 1.5~K and 5~K. No loss of asymmetry is observed, confirming the paramagnetic behavior observed in susceptibility. (c) Magnetic susceptibility in $x = 0.60$ samples is shown as a function of temperature, where the 1~T applied magnetic field is parallel and perpendicular [001]. The anomaly at 68~K corresponds to the ordering temperature ($T_N$). Magnetization shows linear field-dependence up to 10~T both below and above the transition temperature, as seen in the inset. (d) In the weak transverse field $\mu$SR measurements, loss of muon polarization is observed at 70~K and below, suggesting long-range magnetic ordering. (e) The entire sample is found to be magnetically ordered below 50~K, with the magnetic transition occurring over a broad temperature range. The muon decay rate is comparable to the slow (dynamic) component observed in $x = 0.15$ samples. (f) Zero field $\mu$SR data reveal a single-component decay, with no oscillations observed up to 5~K.}
\label{SI MT+muSR others}
\end{figure*}

$\chi(T)$ is shown in Fig. \ref{SI MT+muSR others} (a) for samples with $x = 0.09$ along [001], [010], and [101]. Noticeable anisotropy is exhibited in these samples, and a Curie-Weiss paramagnetic behavior is observed. Curie fit shown in the inset warrants $\mu_{eff} = 0.7 \mu_B$ per f.u.. To probe magnetic properties further, $\mu$SR measurements were performed. No symmetry loss was observed under a weak TF of 3~mT down to 1.5 K, as shown in Fig. \ref{SI MT+muSR others} (b). This finding is consistent with the paramagnetic behavior observed in magnetization measurements. 

For the Cr-rich sample ($x=0.60$), $\chi (T)$ is shown in Fig. \ref{SI MT+muSR others} (c). Curie-Weiss behavior is observed in $\chi (T)$, with discontinuities at 68 K marking the magnetic ordering temperature $T_N$. At $T_N$, a cusp-like feature is observed $\chi(T)$ along [001], while a further increase is observed when $\chi(T)$ is measured perpendicular to [001]. Field-dependent magnetization, measured along [001], shows linear behavior both above and below $T_N$, reaching a maximum of 0.1 $\mu_B$ per f.u at 10~T (see inset of Fig. \ref{SI MT+muSR others} (c)). The plateau in $\chi(T)$ and low $M(H)$ supports a spin-compensated structure, consistent with the prior literature reports \cite{Cr-doped_FeSb2}.

$\mu SR$ data collected in the TF configuration are shown in Fig. \ref{SI MT+muSR others} (d) for representative temperatures. Complete polarization oscillations are visible in the paramagnetic regime at 100~K, while polarization loss is observed at 70~K and below. These spectra were fitted using equation (1) [see Main manuscript] to extract the magnetic volume fraction and decay rates, as shown in Fig. \ref{SI MT+muSR others} (e). Magnetic ordering occurs over a broad temperature range of 60-80~K, below which the entire sample orders magnetically. The decay rate exhibits a temperature-dependent peak well below $T_N$, suggesting only the slow (dynamic component) is observed. No fast decay component was observed within the resolution limit of $mu$SR. In ZF, the slow component is detected down to 5 K [see Fig. \ref{SI MT+muSR others} (f)]. No time-domain oscillations were observed in the ZF spectra, consistent with a non-uniform field distribution at the muon sites.

\pagebreak
\clearpage